\documentclass[twoside,fleqn]{article}
\usepackage[dvips]{graphics}
\usepackage{espcrc2}
\usepackage{epsfig}
\usepackage{latexsym}

\newcommand{\are}{adjoint representation}

\newcommand{\bd}{\beta_{\rm deconf}}
\newcommand{\bc}{\beta_{\rm chiral}}
\newcommand{\td}{T_{\rm deconf}}
\newcommand{\tc}{T_{\rm chiral}}

\title{SU(3) Gauge Theory with Adjoint Fermions}

\author{
  F. Karsch, M. L\"utgemeier
  \thanks{This work was partly supported by the TMR network
    {\it Finite Temperature Phase Transitions in Particle Physics}, EU
    contract no. ERBFMRX-CT97-0122.
    }\\
  Faculty of Physics, University of Bielefeld,
  P.O. Box 100131, 33501 Bielefeld, Germany\\
  }

\begin{document}

\pagestyle{empty}

\begin{abstract}
We analyze the finite temperature phase diagram of QCD with fermions in
the adjoint representation. The simulations performed with four dynamical
Majorana fermions, which is equivalent to two Dirac fermions, show
that the deconfinement and chiral phase transitions occur at two distinct
temperatures,
$T_{\rm chiral} \simeq 6.65 T_{\rm deconf}$. While the deconfinement
transition is first order we find evidence for a continuous chiral
transition. We also present potentials for $T<T_{\rm deconf}$ and
$T_{\rm deconf}<T< T_{\rm chiral}$ both for fundamental and adjoint
fermion-antifermion pairs.
\end{abstract}

\maketitle

\section{Motivation}
In QCD the interplay between confinement and chiral symmetry restoration
is one of the most puzzling problems.
As the two phenomena origin from different non-perturbative
mechanisms it has been speculated that QCD would undergo two distinct
phase transitions \cite{shuryak}.
However, so far all lattice calculations have shown that
there is just one unique critical temperature,
and the critical behaviour close to $T_c$ is influenced by
chiral as well as confinement properties of the system.

Thus, it is of interest to study an $SU(N)$ gauge theory with quarks in
the adjoint representation, where the two transitions fall apart.
This theory has a global $Z(N)$ center symmetry and an $SU(2 n_f)$ chiral
symmetry \cite{peskin}.
Investigations of $SU(2)$ \cite{kogut85,kogut87}
have shown indeed, that the two transitions are well separated with
$\td<\tc$.

In this work we simulate an SU(3) gauge theory with the usual plaquette action
and 2 dynamical staggered (Dirac) fermions in the {\are} 
on $8^3\times 4$ and $16^3\times 4$ lattices
using the exact hybrid $\Phi$ pseudo fermion algorithm.
We ran at up to 15 $\beta$ values in the interval $[5.2;7.0]$ with 
a mass between 0.10 and 0.01 (small lattice) resp. $m=0.02$ (larger lattice).

We note that our model is closely related to
super-symmetric gauge theories (see \cite{witten}).

\section{Phase Structure}
In order to clarify the phase structure of our model we analyze the
Polyakov loop and the chiral condensate, which are the order parameters
for [de-]confinement resp. the chiral transition.
On the smaller lattice, which we only discuss in this section, we simulated
with several mass values allowing us to extrapolate to the zero mass case.

\begin{figure}[hbt]
  \setlength{\unitlength}{0.80mm}

  \begin{picture}( 90, 62)(0,0)
    \put(  0.0,  0.0){\epsfig{file=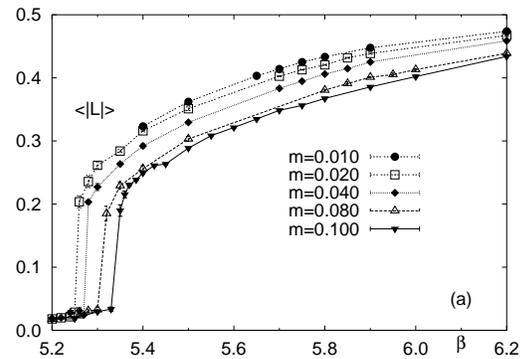,width=90\unitlength}}
  \end{picture}

  \vspace*{-7ex}
  \caption{Polyakov loop for several mass and
    $\beta$ values on the $8^3\times 4$ lattice}
  \label{fig:poly}
\end{figure}

\paragraph{Polyakov loop}
In figure \ref{fig:poly} one sees that the Polyakov loop
shows a first order signal around $\beta=5.3$, which is confirmed by
a double peak structure of the combined plaquette histogram

\begin{figure*}[t]
  \setlength{\unitlength}{0.80mm}
  \begin{picture}( 90, 62)(0,0)
    \put(  0.0,  0.0){\epsfig{file=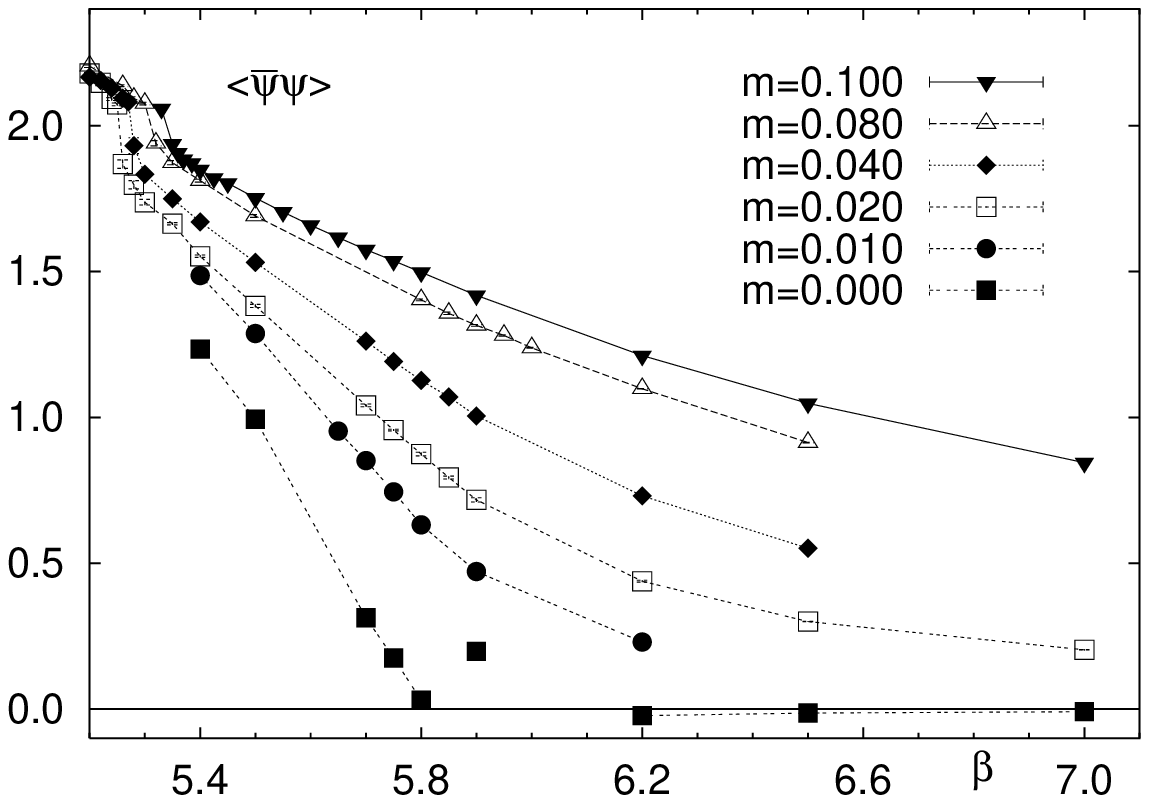,width=90\unitlength}}

    \put( 16.6, 15.0){\small (a)}
  \end{picture}
  \hspace*{\fill}
  \begin{picture}( 90, 62)(0,0)
    \put(  0.0,  0.0){\epsfig{file=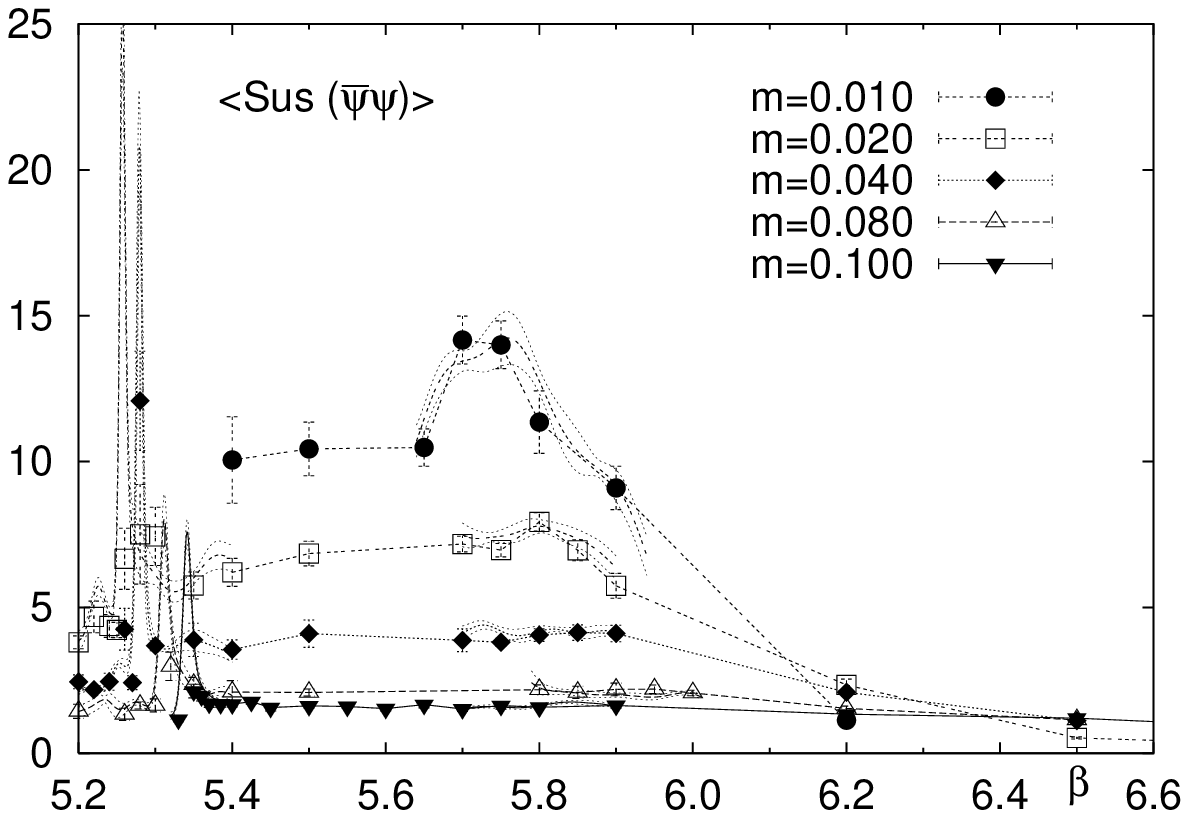,width=90\unitlength}}

    \put( 78.0, 15.0){\small (b)}
  \end{picture}
  \hspace*{\fill}

  \vspace*{-7ex}
  \caption{Chiral condensate (a) with $m\to 0$ extrapolation (filled squares)
    and its susceptibility (b) on the $8^3\times 4$ lattice}
  \label{fig:chisch}
\end{figure*}

In the following table the critical couplings are summarized, which are
obtained from the re-weighted Polyakov loop susceptibilities.

\centerline{\tabcolsep0.5ex
\begin{tabular}[t]{|c|c|c|c|c|}
  \hline
  mass  \rule[-1ex]{0em}{3.4ex} & 0.10     & 0.08     & 0.04     & 0.02    \\
  \hline
  $\bd$ \rule[-1ex]{0em}{3.4ex} & 5.342(2) & 5.312(2) & 5.279(2) & 5.256(2)\\
  \hline
\end{tabular}}
\vspace{1.5ex}

They can be fitted to a linear function with $\beta_{\rm critical} = 5.236 (3)$
for $m=0$.


\paragraph{Chiral condensate}
For the chiral condensate we must go to the zero mass limit to find the
critical coupling, where it drops to zero.
This extrapolation of $\bar{\psi}\psi$ to $m=0$ is done with the following
fit ansatz.

\vspace{-0.5ex}
\begin{equation}
  \label{eq:fitansatz}
  \langle \bar{\psi}\psi \rangle (\beta,m)
  = a_0 + \bar{a}_1 m^{1/\delta} + a_1 m + a_2 m^2 + \ldots
\end{equation}
\vspace{-0.5ex}

For $\beta\le 5.8$ a nonzero $a_0$ and a leading $\sqrt{m}$-term is found,
above we find a polynomial with $a_0\approx 0$.
In figure \ref{fig:chisch}a the finite mass data for $\bar{\psi}\psi$
and the extrapolated curve is shown.
From the latter we can read of $\bc\approx 5.8$, which we can check by looking
at the re-weighted chiral susceptibility (fig. \ref{fig:chisch}b).

The first peak around $\beta=5.3$ coincides with the deconfinement transition,
then in $[5.4,5.7]$ the chiral susceptibility stays constant. The values of
this plateau is mass dependent and consistent with
$1/\sqrt{m}$, the $m$-derivative of equation \ref{eq:fitansatz}.

Around $\beta=5.8$ there is (for $m\le0.02$) a second peak indicating
the chiral phase transition.
For $m=0.02$ we get $\Delta\beta=0.544(32)$ and using the two-loop
$\beta$-function we find $\tc/\td \approx 6.65 \pm 1.13$.
In $SU(2)$ a qualitative similar behaviour was found with a ratio of $175 \pm
50$ \cite{kogut85}.

The analysis of the larger lattice at $m=0.02$ confirms our observations, large
finite size effects seem to be absent.
In summary our model contains a mixed phase (roughly in $[5.3;5.8]$),
where the quarks are already deconfined but with the chiral symmetry still
broken.


\section{Potentials from Polyakov Loop Correlations}
\begin{figure*}[htbp]
  \setlength{\unitlength}{0.80mm}

  \begin{picture}(90,62)(0,0)
    \put(  0.0,  0.0){\epsfig{file=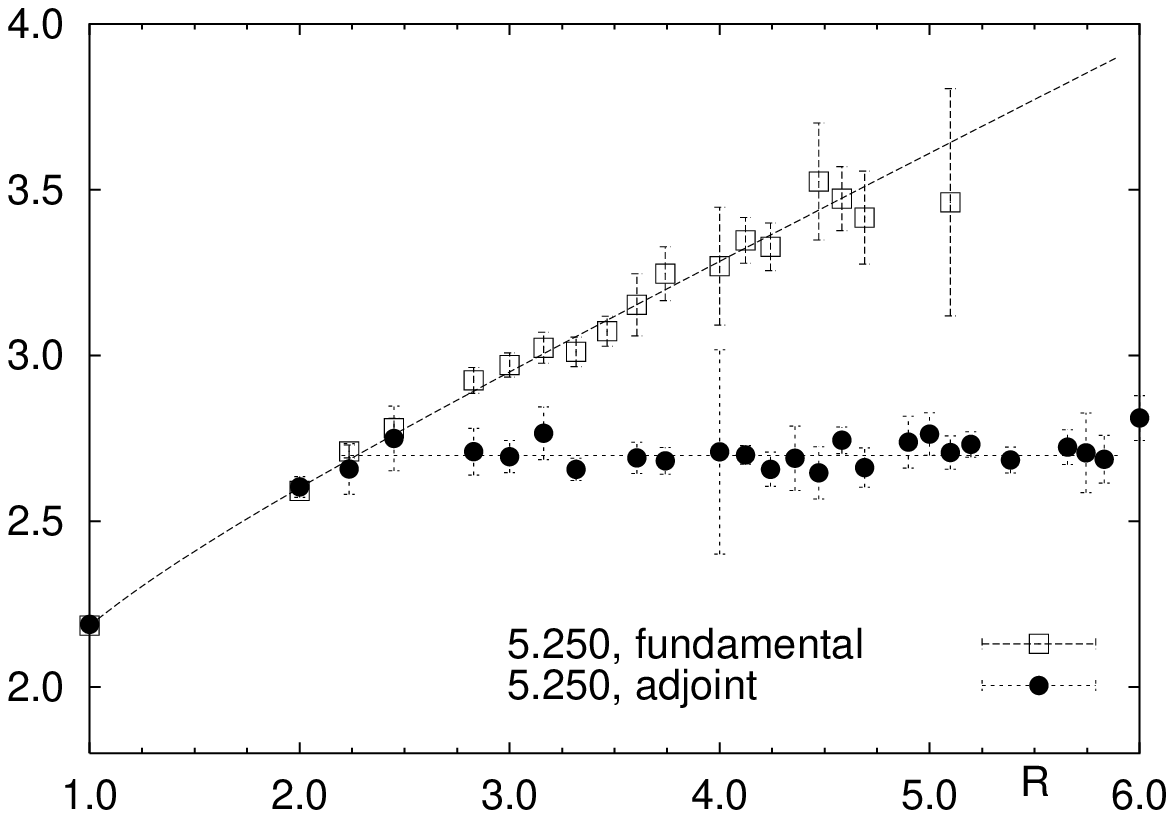,width=90\unitlength}}

    \put( 25.0, 45.0){\small $V(R)/T$}
    \put( 16.6, 36.0){\small (a)}
  \end{picture}
  \hspace*{\fill}
  \begin{picture}(90,62)(0,0)
    \put(  0.0,  0.0){\epsfig{file=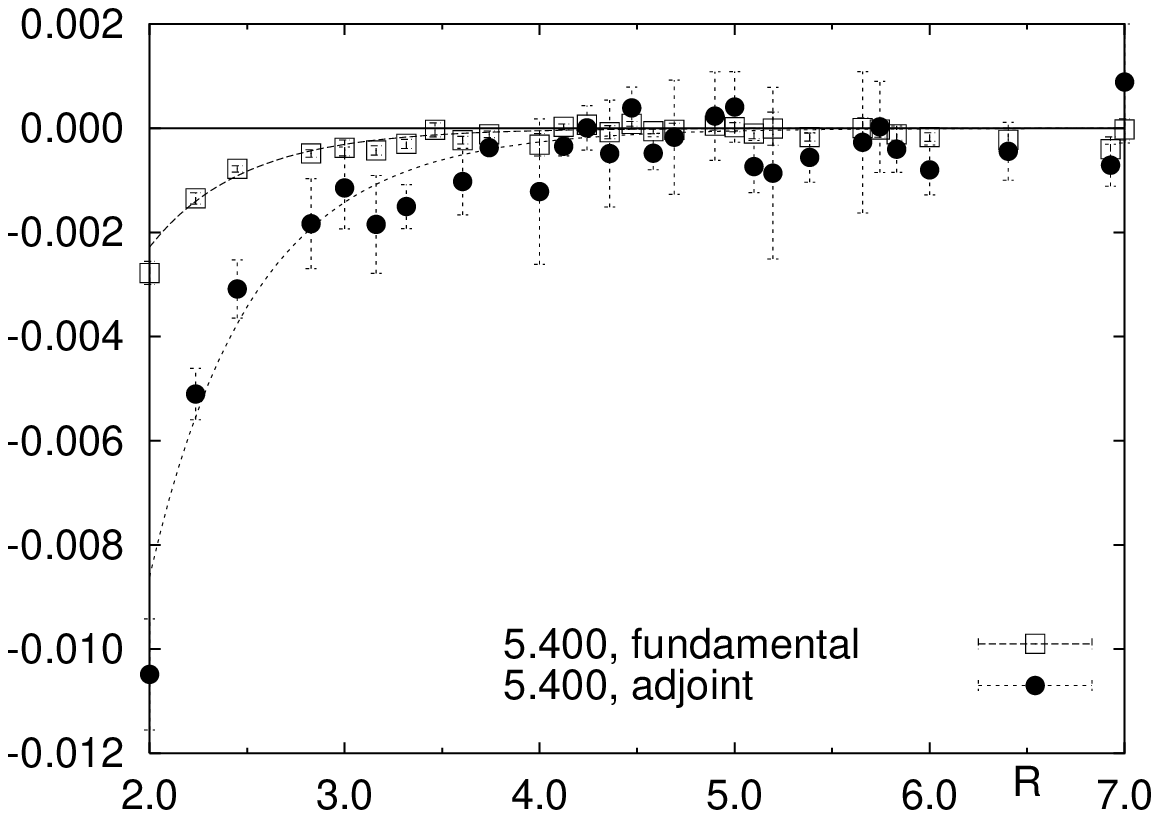,width=90\unitlength}}

    \put( 25.0, 30.0){\small $V(R)/T$}
    \put( 78.0, 36.0){\small (b)}
  \end{picture}
  \hspace*{\fill}

  \vspace*{-7ex}
  \caption{Potentials in the confined phase (a) and the mixed phase (b) on the
    $16^3\times 4$ lattice}
  \label{fig:potbelow}
\end{figure*}

Correlations of fundamental as well as adjoint Polyakov loops are used to get
the potential of a static quark-antiquark pair. The potentials for two
couplings ($5.25, 5.40$) are shown in figure \ref{fig:potbelow}.

At $\beta=5.25$, i.e. below the deconfinement transition, the
fundamental potential can be fitted to

\vspace{-0.5ex}
\begin{equation}
  \label{eq:potfit1}
  V(R) = V_0 + \sigma R - \alpha/R
\end{equation}
\vspace{-0.5ex}

with $\sigma a^2=0.317(7)$ and $\alpha=0.198(5)$
while the adjoint potential shows a string-breaking behaviour. At distances
larger than $2.2$ it is consistent with the constant $V_\infty a=0.270(1)$.

At $\beta=5.40$ both (normalized) potentials can be fitted to

\vspace{-0.5ex}
\begin{equation}
  \label{eq:potfit2}
  V(R) = - \alpha/R \exp (-\mu R)
\end{equation}
\vspace{-0.5ex}

with these results:

\vspace{-0.25ex}
\centerline{
\begin{tabular}[t]{|c|c|c|c|}
  \hline
  ~            \rule[-1ex]{0em}{3.4ex} & $\alpha$  & $\mu a$  \\
  \hline
  fundamental  \rule[-1ex]{0em}{3.4ex} & 0.11 (4) & 1.59 (15) \\
  \hline
  adjoint      \rule[-1ex]{0em}{3.4ex} & 0.28 (9) & 1.39 (14) \\
  \hline
\end{tabular}}


\section{Thermodynamics around $\td$}
To calculate the pressure the $\beta$-function and plaquette data for
$T=0$ are needed (\cite{boyd,engels}).
Without it we can only speculate about the general
behaviour of the pressure.

The plaquette data show a sharp rise at $\bd$ but
behave moderately around $\bc$ (see figures in \cite{mluetgem}).
Therefore the pressure (the integrated plaquette) will also rise at
$\td$ but will probably not feel the chiral symmetry restoration.          
The latent heat can be derived from the gap in the plaquette and the
$\beta$-function.
The former is estimated from the re-weighted plaquette while the latter
is known here only perturbatively.

\vspace{-0.8ex}
\begin{equation}
  \label{eq:latentheat1}
  \frac{\Delta\epsilon}{T^4} = \frac{\Delta(\epsilon-3p)}{T^4} 
  = - a \frac{d\beta}{da} \cdot N_{\tau}^4 \cdot 6 \, \Delta \mbox{Plaq}
\end{equation}
\vspace{-0.8ex}

\begin{figure}[htbp]
  \setlength{\unitlength}{0.60mm}

  \begin{picture}(67,62)(0,0)
    \put( -5.0,  0.0){\epsfig{file=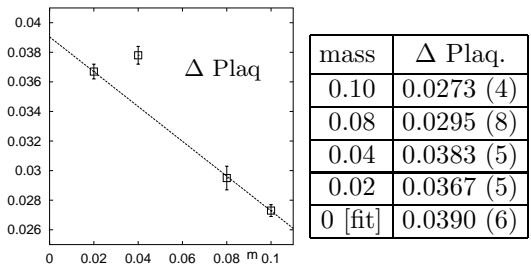,width=72\unitlength}}
    \put( 41.0, 45.0){\small $\Delta$ Plaq}
  \end{picture}
  \parbox[b]{29mm}{
    \tabcolsep0.8ex
    \begin{tabular}{|c|c|}
      \hline
      mass \rule[-1ex]{0em}{3.4ex} & $\Delta$ Plaq. \\ \hline
      0.10    & 0.0273 (4) \\ \hline
      0.08    & 0.0295 (8) \\ \hline
      0.04    & 0.0383 (5) \\ \hline
      0.02    & 0.0367 (5) \\ \hline
      0 [fit] & 0.0390 (6) \\ \hline
    \end{tabular}

    \vspace*{3.5ex}
    }

  \vspace*{-7ex}
  \caption{Estimated plaquette differences at $\bd$ and their extrapolation to $m=0$}
  \label{fig:final}
\end{figure}

Results are shown in figure \ref{fig:final}.
For $m=0$ we finally estimate

\vspace{-0.8ex}
\begin{equation}
  \label{eq:latentheat2}
  \frac{\Delta\epsilon}{T^4}
  = 14.4 = 0.36 \,\cdot\, \frac{\epsilon_{SB}(N_{\tau}=4)}{T^4}
\end{equation}
\vspace{-0.8ex}

This result is very close to the pure gauge case, where the latent
heat is about 34\% of the Stefan-Boltzmann limit at $N_{\tau}=4$
\cite{iwasaki}. On the other hand the latent heat per degree of
freedom in the deconfined phase is much lower, because there are now
64 fermionic degrees of freedom instead of 24.

\section{Conclusions}
\sloppy
Analyzing $SU(3)$ with 2 adjoint fermions
we find two distinct phase transitions with $\tc/\td\approx 6.65$.
In the intermediate phase (deconfined but chirally broken)
the chiral condensate depends on $\sqrt{m}$.
The deconfinement transition seems to be first order with a latent
heat of 35\% of the Stefan-Boltzmann limit while the chiral transition is
continuous.
We also note that it may be interesting to study this model at non-zero
chemical potential as its fermion determinant stays real and positive in this
case.

\end{document}